\documentclass[aps,showpacs,twocolumn,superscriptaddress]{revtex4}
\usepackage[dvips]{graphicx}
\usepackage{subfigure}
\usepackage{amsmath}
\usepackage{color}

\begin{document}

\title{Temperature-dependent thermal conductivities of one-dimensional nonlinear Klein-Gordon lattices with soft on-site potential}

\author{Linlin Yang}
\affiliation{Center for Phononics and
Thermal Energy Science and School of Physics Science and
Engineering, Tongji University, 200092 Shanghai, People's Republic
of China}

\author{Nianbei Li}
\email{nbli@tongji.edu.cn} \affiliation{Center for Phononics and
Thermal Energy Science and School of Physics Science and
Engineering, Tongji University, 200092 Shanghai, People's Republic
of China}

\author{Baowen Li}
\email{phononics@tongji.edu.cn} \affiliation{Center for Phononics
and Thermal Energy Science and School of Physics Science and
Engineering, Tongji University, 200092 Shanghai, People's Republic
of China} \affiliation{Department of Physics and Centre for
Computational Science and Engineering, National University of
Singapore, Singapore 117546, Republic of Singapore} \affiliation{NUS
Graduate School for Integrative Sciences and Engineering, Singapore
117456, Republic of Singapore}

\begin{abstract}
The temperature-dependent thermal conductivities of one-dimensional nonlinear Klein-Gordon lattices with soft on-site potential (soft-KG) have been investigated systematically. Similar to the previously studied hard-KG lattices, the existence of renormalized phonons has also been confirmed in soft-KG lattices. In particular, the temperature-dependence of renormalized phonon frequency predicted by a classical field theory has been verified by detailed numerical simulations. However, the thermal conductivities of soft-KG lattices exhibit opposite trend in the temperature dependence in comparison with the hard-KG lattices. The interesting thing is that both the temperature-dependent thermal conductivities of soft- and hard-KG lattices can be interpreted in the same framework of effective phonon theory. According to the effective phonon theory, the exponents of the power-law dependence of the thermal conductivities as the function of temperature are only determined by the exponents of the soft or hard on-site potentials. These theoretical predictions have been consistently verified very well by extensive numerical simulations.
\end{abstract}

\maketitle
\section{Introduction}
Since the first discovery of anomalous heat conduction in 1D Fermi-Pasta-Ulam $\beta$ (FPU-$\beta$) lattices where thermal conductivity diverges with the lattice size~\cite{Lepri1997prl}, enormous efforts have been put forward to the study of heat transport in 1D nonintegrable lattices trying to unravel the underlying physical mechanism~\cite{Lepri1998epl,Hu1998pre,Lepri1998pre,Tong1999prb,Hatano1999pre,Tsironis1999pre,Sarmiento1999pre,Dhar1999prl,Alonso1999prl,Hu2000pre,Aoki2000pla,Li2001prl,
Dhar2001prl,Aoki2001prl,Zhang2002pre,Li2002prl,Narayan2002prl,Saito2003epl,Savin2003pre,Lepri2003pre,Segal2003jcp,Wang2004prl,Gendelman2004prl,Li2005chaos,
Cipriani2005prl,Zhang2005jcp,Zhao2006prl,Chang2008prl,Dubi2009pre,Henry2009prb,Saito2010prl,Wang2010prl,Yang2010nt,Wang2011epl,Wang2012pre,Landi2013pre,Mendl2013prl,Xu2014nc,Liu2014prl,Meier2014prl}. The consensus reached in this community is that the total momentum conservation do play an important role in determining the system's heat conduction behavior. For momentum-nonconserving nonintegrable lattices with on-site potentials, there is no dispute that they should have normal heat conduction. However, for momentum-conserving nonintegrable lattices, the issue whether they should necessarily give rise to anomalous heat conduction is still under severe debate. The hydrodynamic theory has predicted that momentum conservation will naturally induce anomalous heat conduction in 1D nonintegrable lattices~\cite{Narayan2002prl}. But this theory fails to explain the normal heat conduction numerically found for the rotor model which is also a momentum conserving lattice~\cite{Gendelman2000prl,Giardina2000prl,Li2014arxiv}. There are some arguments that the 1D rotor model should exhibit anomalous heat conduction in thermodynamical limit where approval or disapproval is very hard to be performed by numerical simulations due to huge computation cost~\cite{Flach2003chaos}. Most recently, the normal or anomalous heat transport in asymmetrical momentum conserving lattices has attracted much attention~\cite{Zhong2012pre,Wang2013pre,Das2014jsp}, but more studies in this new issue need to be done when a final conclusion can be drawn. For a thorough comprehension of the heat transport problems in low dimensional systems, please refer to some excellent review articles for further reading~\cite{Lepri2003pr,Dhar2008ap,Liu2013epjb}.

Despite that the study of normal/anomalous heat conduction or size-dependent thermal conductivities is still causing many troubles for this community, the exploration of temperature dependence of thermal conductivities in 1D nonintegrable lattices turns out to be more successful. The difficulty of study on size dependence arises from the fact that numerical simulations on extremely long lattices need to be calculated in order to reach the asymptotic behavior. In contrast, a very short lattice usually in the number of hundreds or thousands atoms is enough to obtain stable temperature dependence of thermal conductivities~\cite{Aoki2000pla,Aoki2001prl,Li2007epl,Li2007pre,He2008pre,Shao2008pre,Li2009jpsj,Li2012aip,Li2013pre}. The most important thing is that the diversity of temperature dependence for various lattice models can provide perfect testbed for any heat conduction theory, especially some kind of phenomenological theory. According to numerical simulations, the thermal conductivities of FPU-$\beta$ lattice depend on temperature as $\kappa(T)\propto 1/T$ in the low temperature limit and $\kappa(T) \propto T^{1/4}$ in the high temperature limit~\cite{Aoki2001prl,Li2007epl,Li2012aip}. The $H_n$ models have monotonically ascending temperature dependence for thermal conductivities as $\kappa(T)\propto T^{1/2-1/n}$~\cite{Li2012aip}, while the hard-KG lattices exhibit monotonically descending temperature dependence as $\kappa(T)\propto T^{-\frac{4(n-2)}{n+2}}$ where $n>2$ is the exponent of the on-site potential~\cite{Li2013pre}. Interesting enough, all these temperature dependence can be {\it quantitatively} and {\it consistently} interpreted by the same phenomenological effective phonon theory~\cite{Li2007epl,Li2012aip,Li2013pre} which is based on the renormalized phonons~\cite{Alabiso1995jsp,Lepri1998pre,Alabiso2001jpa,Boyanovsky2004prd,Gershgorin2005prl,Li2006epl,Gershgorin2007pre,Li2010prl} occurred in nonlinear lattices.

In this work, we would like to extend the analytical and numerical investigations of temperature dependence of thermal conductivities to several soft-KG lattices with $1<n<2$. The spectra of renormalized phonons will be numerically calculated in comparison to the predictions from classical field theory. The temperature-dependent thermal conductivities will be systematically studied via non-equilibrium molecular dynamics. Consistent and quantitative comparison between numerical simulations and theoretical predictions from effective phonon theory will be performed. The paper is organized as the follows: in Sec. II the soft-KG lattice models will be introduced and the properties of renormalized phonons will be presented. Sec. III will then display the results of temperature-dependent thermal conductivities and their comparison with the theoretical predictions from effective phonon theory. We will give conclusions and summaries in Sec. IV.

\begin{figure}
\subfigure[~Soft-KG lattice with $n=1.25$.]{\includegraphics[width=0.8\columnwidth]{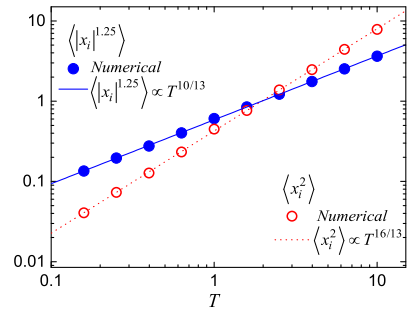}}
\subfigure[~Soft-KG lattice with $n=1.50$.]{\includegraphics[width=0.8\columnwidth]{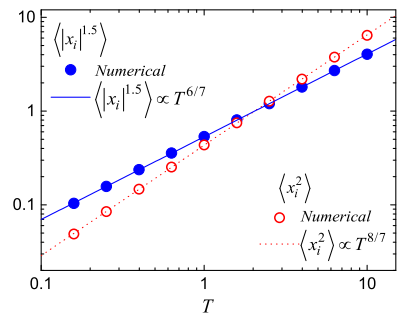}}
\subfigure[~Soft-KG lattice with $n=1.75$.]{\includegraphics[width=0.8\columnwidth]{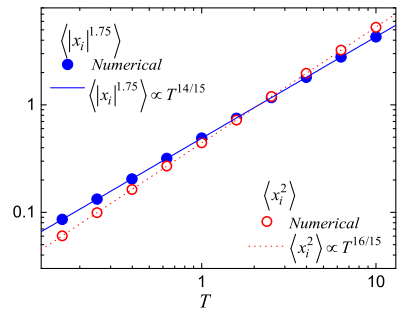}}
\caption{\label{fig:scaling-phi4} (color online).
The time averages of $\left<x^2_i\right>$ and $\left<\left|x_i\right|^n\right>$
as the function of temperature $T$ for three soft-KG lattices with (a) $n=1.25$; (b) $n=1.5$; (c) $n=1.75$. The lines are guided for the eyes.
All the numerical simulations are performed at thermal equilibrium for
a lattice with $N=1600$ where the two ends are coupled to the Langevin
heat baths. The averages of $\left<x^2_i\right>$ and $\left<\left|x_i\right|^n\right>$
are independent of atom index $i$.}\label{fig:scaling-x24}
\end{figure}

\section{The soft-KG lattices and their renormalized phonons}

The symmetrical nonlinear KG lattices have the following Hamiltonian:
\begin{equation}\label{KG}
H=\sum^{N}_{i=1}\left[\frac{1}{2}p^2_i+\frac{1}{2}(x_i-x_{i-1})^2+\frac{1}{n}\left|x_i\right|^n\right]
\end{equation}
where $x_i$ and $p_i$ denote the dimensionless displacement and momentum for $i$-th atom and $n$ is the exponent of nonlinear on-site potential. For $n>2$, the on-site potentials are hard types which are harder than the referenced quadratic potential with $n=2$. The Hamiltonian of Eq. (\ref{KG}) with $1<n<2$ is then called soft-KG lattices. In contrast to hard-KG lattices, the soft-KG lattices approach to the harmonic system in the high temperature limit. In order to get consistent understanding for the thermal properties of soft-KG lattices, three different soft-KG lattices with $n=1.25,1.50$ and $1.75$ will be investigated systematically. The dimensionless units have been applied. For simplicity, the periodic boundary conditions with $x_i=x_{N+i}$ will be used for theoretical analysis while fixed boundary conditions with $x_0=x_{N+1}=0$ will be used for molecular dynamics simulations. In principle, the different boundary conditions will not cause any difference for their thermal properties in thermodynamical limit.

The dispersion relation of renormalized phonons for Hamiltonian of Eq. (\ref{KG}) can be generally expressed as~\cite{Boyanovsky2004prd,Li2013pre}:
\begin{equation}\label{spectrum}
\hat{\omega}_k=\sqrt{\omega^2_k+\gamma},\,\,\omega_k=2\sin{\frac{\pi k}{N}},\,\,\gamma=\frac{\sum_i\left<\left|x_i\right|^n\right>}{\sum_i\left<x^2_i\right>}
\end{equation}
where $k,i=1,...,N$ and $\left<\cdot\right>$ denotes the ensemble average at thermal equilibrium. The renormalization coefficient $\gamma$ contains the information of nonlinearity and depends on the temperature or the strength of nonlinearity. It is very interesting that $\gamma$ depends on temperature with a power law behavior which can be predicted by the classical field theory approach~\cite{Boyanovsky2004prd,Li2013pre}.

\begin{figure}
\includegraphics[width=\columnwidth]{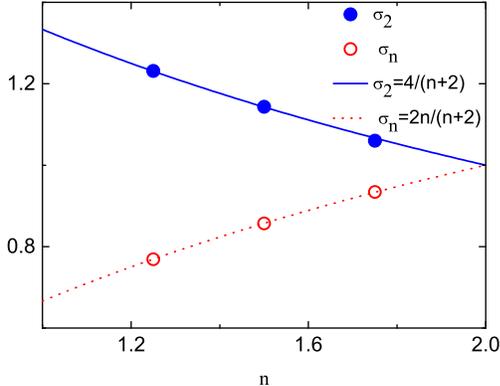}
\vspace{-0.5cm} \caption{\label{fig:sigma-phin} (color online). The
exponents $\sigma_2$ and $\sigma_n$ as the function of $n$. The
symbols are numerical data and the lines are predictions of Eq.
(\ref{scaling-x2n}).}\label{fig:sigma-2n}
\end{figure}

According to Ref. \cite{Li2013pre}, the scaling of components $\left<x^2_i\right>$ and $\left<\left|x_i\right|^n\right>$ of $\gamma$ can be derived as the function of temperature as following:
\begin{eqnarray}\label{scaling-x2n}
\left<x^2_i\right>&\propto& T^{\sigma_2},\,\,\sigma_2=\frac{4}{n+2}\nonumber\\
\left<\left|x_i\right|^n\right>&\propto& T^{\sigma_n},\,\,\sigma_n=\frac{2n}{n+2}
\end{eqnarray}
where $\left<x^2_i\right>$ and $\left<\left|x_i\right|^n\right>$ are independent of the atom index $i$. For soft-KG lattice with $n=1.25$, the power-law dependence can be expressed as $\left<x^2_i\right>\propto T^{16/13}$ and $\left<\left|x_i\right|^n\right>\propto T^{10/13}$. Similarly, the expressions are $\left<x^2_i\right>\propto T^{8/7}$ and $\left<\left|x_i\right|^n\right>\propto T^{6/7}$ for $n=1.5$ and $\left<x^2_i\right>\propto T^{16/15}$ and $\left<\left|x_i\right|^n\right>\propto T^{14/15}$ for $n=1.75$. To verify these theoretical predictions from classical field theory, we have numerically calculated the time average of $\left<x^2_i\right>$ and $\left<\left|x_i\right|^n\right>$ as the function of temperature for three different soft-KG lattices with exponents $n=1.25,1.5,1.75$ as plotted in Fig. \ref{fig:scaling-x24}. It can be seen clearly that all the $\left<x^2_i\right>$ and $\left<\left|x_i\right|^n\right>$ follow the predicted power-law dependence as the function of temperature over two order of magnitudes. In Fig. \ref{fig:sigma-2n}, the fitting exponents $\sigma_2$ and $\sigma_n$ extracted from Fig. \ref{fig:scaling-x24} have been plotted compared with the theoretical predictions of Eq. (\ref{scaling-x2n}). The good agreement between numerical simulations and theoretical predictions can be easily found.

\begin{figure}
\subfigure[~Soft-KG lattice with $n=1.25$.]{\includegraphics[width=0.85\columnwidth]{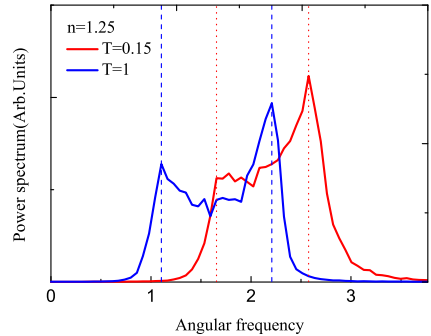}}
\subfigure[~Soft-KG lattice with $n=1.50$.]{\includegraphics[width=0.85\columnwidth]{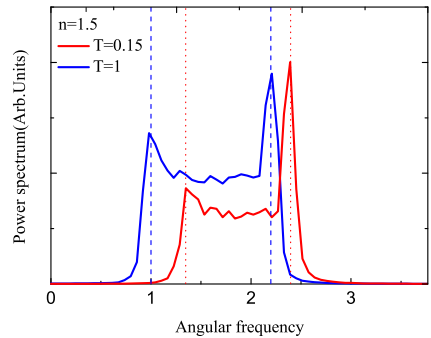}}
\subfigure[~Soft-KG lattice with $n=1.75$.]{\includegraphics[width=0.85\columnwidth]{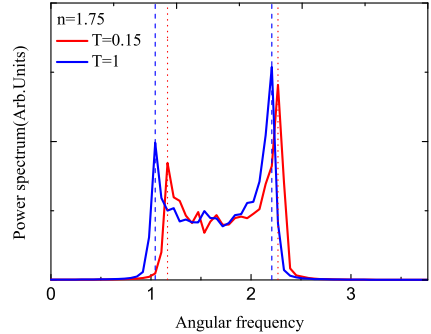}}
\caption{\label{fig:ps} (color online).
The power spectra of atom velocity $\dot{x}_i(t)$ at two different temperatures $T=0.15$ and $T=1$ for three soft-KG lattices with (a) $n=1.25$; (b) $n=1.5$; (c) $n=1.75$. The predicted boundaries of power spectra are marked by vertical red dotted lines for $T=0.15$ and blue dashed lines for $T=1$, respectively. All the numerical simulations are performed at thermal equilibrium for
a lattice with $N=200$.}\label{fig:ps}
\end{figure}

In order to confirm the existence of the renormalized phonons predicted by Eq. (\ref{spectrum}), we should investigate the power spectrum of atom vibrations in different soft-KG lattices at different temperatures. From the above discussion of Eq. (\ref{spectrum}) and (\ref{scaling-x2n}), we know that the renormalization coefficient $\gamma$ can be expressed as
\begin{equation}\label{T-gamma}
\gamma=\xi\cdot T^{\sigma_n/\sigma_2}
\end{equation}
where the prefactor $\xi$ is a temperature-independent constant. In lack of theoretical predictions, the prefactor $\xi$ should be obtained from numerical simulations for each soft-KG lattice. For $n=1.25,1.5$ and $1.75$, the numerical values of $\xi$ have been calculated as $1.377,1.016$ and $1.107$, respectively. According to Eq. (\ref{spectrum}), the renormalized phonon spectra are bounded as
\begin{equation}\label{sp-region}
\hat{\omega}_k \in \left[\sqrt{\xi\cdot T^{\sigma_n/\sigma_2}}, \,\,\,\sqrt{4+\xi\cdot T^{\sigma_n/\sigma_2}}\right]
\end{equation}
where the upper and lower boundaries are both temperature dependent. The spectra of predicted renormalized phonons are therefore located within $[1.656,2.576]$ at $T=0.15$ and $[1.104,2.208]$ at $T=1$ for lattice with $n=1.25$. For $n=1.5$, the predicted region is $[1.347,2.397]$ at $T=0.15$ and $[1.0,2.198]$ at $T=1$. For $n=1.75$, the region should be $[1.165,2.269]$ and $[1.043,2.208]$ for $T=0.15$ and $T=1$, respectively.

The phonon spectra can be obtained by calculating the power spectra of atom velocity $\dot{x}_i(t)$ for each soft-KG lattice at specified temperature~\cite{Li2012rmp}. In Fig. \ref{fig:ps}, the power spectra for soft-KG lattices at different temperatures have been plotted. The vertical lines are the predictions from Eq. (\ref{sp-region}) which exactly match the numerical boundaries. The good agreement between numerical results and theoretical predictions confirms the existence of renormalized phonons in these nonlinear soft-KG lattices.

\section{Temperature-dependent thermal conductivities and effective phonon theory}
In dealing with the temperature-dependent thermal conductivities, the effective phonon theory~\cite{Li2006epl,Li2007epl,Li2012aip,Li2013pre} has been proved to be very successful. In the framework of renormalized phonons, the effective phonon theory is able to predict the temperature dependence of thermal conductivities. The derivation of the thermal conductivities for the 1D soft-KG lattices is same as that for 1D hard-KG lattices. In particular, the thermal conductivities of 1D soft-KG can be expressed as~\cite{Li2013pre}
\begin{equation}\label{ept}
\kappa(T)\propto \frac{1}{\epsilon\gamma^{3/2}}
\end{equation}
where $\gamma$ is the renormalization coefficient and the nonlinearity strength $\epsilon$ is defined as the ratio between nonlinear potential energy and total potential energy: $\epsilon=\left<E_n\right>/\left<E_t\right>$ with $E_n$ and $E_t$ denoting the nonlinear and total potential energy, respectively.

the temperature dependence of $\gamma$ has been derived in Eq. (\ref{T-gamma}). Next we will briefly introduce the derivation of the temperature dependence of the nonlinearity strength $\epsilon$. From definition, the nonlinearity strength $\epsilon$ can be expressed as
\begin{eqnarray}\label{T-epsilon}
\epsilon&=&\frac{\sum_i\left<\left|x_i\right|^n\right>/n}{\sum_i\left<(x_i-x_{i-1})^2\right>/2+\sum_i\left<\left|x_i\right|^n\right>/n}\nonumber\\
&\approx&\frac{\sum_i\left<\left|x_i\right|^n\right>/n}{\sum_i\left<(x_i-x_{i-1})^2\right>/2}\nonumber\\
&\propto&T^{\sigma_n-1}
\end{eqnarray}
Since we only consider the high temperature region where the quadratic interaction potential $\sum_i\left<(x_i-x_{i-1})^2\right>$ dominates, the on-site potential term $\sum_i\left<\left|x_i\right|^n\right>$ in the denominator can be ignored. From Eq. (\ref{scaling-x2n}) we know that  $\left<\left|x_i\right|^n\right>\propto T^{\sigma_n}$. And the property of $\left<(x_i-x_{i-1})^2\right>\propto T$ can be obtained due to equipartition theorem~\cite{Li2013pre} in the high temperature limit.

Therefore, from Eq. (\ref{ept}), (\ref{T-gamma}) and (\ref{T-epsilon}), the thermal conductivities for 1D soft-KG lattices can be derived as:
\begin{equation}\label{kappa-T-phin}
\kappa(T)\propto T^{r_n},\,\,\,r_n=\frac{4(2-n)}{n+2}
\end{equation}
This power law dependence is exactly the same as that for hard-KG lattices~\cite{Li2013pre}. In both soft- and hard-KG lattices, the considered temperature regions are all close to the harmonic limit. One should notice that this region refers to low temperature region for hard-KG lattices and high temperature region for soft-KG lattices. Our theory predicts that the thermal conductivities increase monotonically with temperature as $r_n>0$ for soft-KG lattices with $n<2$ which is totally different with the hard-KG lattices where $r_n<0$ for $n>2$. To verify these predictions, intensive numerical simulations of temperature-dependent thermal conductivities for soft-KG lattices need to be performed.

\begin{figure}
\includegraphics[width=\columnwidth]{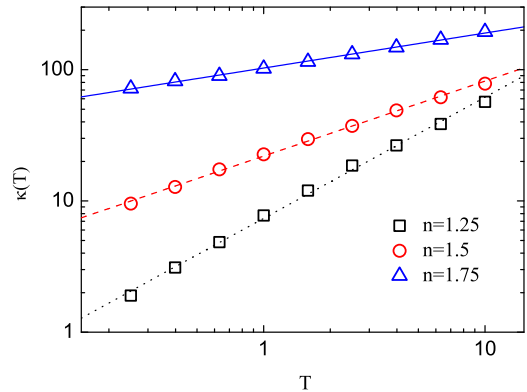}
\vspace{-0.5cm} \caption{\label{fig:kappa-phin} (color online). The
thermal conductivities $\kappa(T)$ as the function of temperature
$T$ for different nonlinear KG lattices with $n=1.25,1.5,1.75$.
The dotted/dashed/solid lines are the numerical fittings of the form $\kappa=A_n
T^{r_n}$ where $A_n$ and $r_n$ are the fitting parameters.}
\end{figure}

In Fig. \ref{fig:kappa-phin}, we numerically calculated the thermal conductivities as the function of temperature for the soft-KG lattices with $n=1.25,1.5$ and $1.75$ from nonequilibrium molecular dynamics simulations. For each thermal conductivity at specific temperature, we have eliminated the size effect by using long enough lattices where the saturation of thermal conductivity has been confirmed. The power law dependence of thermal conductivities as the function of temperature can be easily seen by the straight lines in the log-log scaled plot. In contrary to the hard-KG lattices, the thermal conductivities of all soft-KG lattices increase with temperature monotonically. These are consistent with the prediction of effective phonon theory where $n=2$ in Eq. (\ref{kappa-T-phin}) is the crossover point as $r_2=0$. The thermal conductivities of harmonic KG lattice with $n=2$ must be independent of temperature as we should expect.

The soft-KG and hard-KG lattices display opposite temperature dependence of their thermal conductivities. The physical interpretation behind this scenario is that the nonlinear lattices should possess higher thermal conductivities when they approach further to the harmonic limit. For hard-KG lattices, the harmonic limit is approached for low temperature limit. Therefore the hard-KG lattices have higher thermal conductivities in low temperature region. However, the soft-KG lattices approach to harmonic limit in high temperature regime. Their thermal conductivities thus have higher values in high temperature region.

\begin{figure}
\includegraphics[width=\columnwidth]{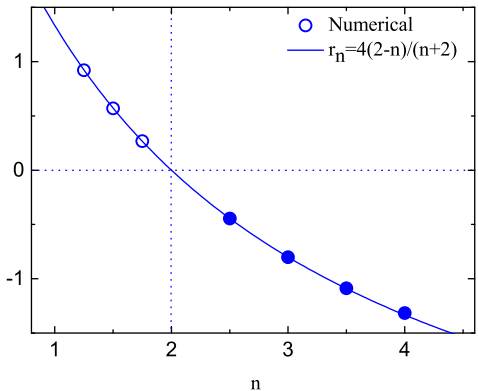}
\vspace{-0.5cm} \caption{\label{fig:nu-phin} (color online). The
exponents $r_n$ as the function of $n$ for $n=1.25,1.5,1.75$ of soft-KG lattices and $n=2.5,3,3.5,4$ of hard-KG lattices.
The hollow circles are numerical data of $r_n$ extracted out from Fig.
\ref{fig:kappa-phin} within the temperature range of $[0.15, 10]$. The solid circles are numerical data referenced from \cite{Li2013pre}. The
dotted line is the theoretical prediction of Eq.
(\ref{kappa-T-phin}) from effective phonon theory.}
\end{figure}

For each soft-KG lattice, a numerical value of the exponent $r_n$ can be extracted out with standard fitting procedure. The resulted exponents $r_n$ are plotted in Fig. \ref{fig:nu-phin} in comparison with the theoretical prediction of Eq. (\ref{kappa-T-phin}). It can be seen that the numerical data are in fully agreement with the prediction of effective phonon theory. Therefore the effective phonon theory, developed from the assumption of renormalized phonons, can also explain {\it quantitatively} and {\it consistently} the temperature dependence of thermal conductivities for soft-KG lattices.

\section{discussion}
In summary, we have systematically studied the thermal properties of soft-KG lattices with nonlinear exponent $n<2$. The renormalized phonons have been confirmed for these soft-KG lattices as the temperature-dependence of renormalization coefficients can be well explained by a classical field theory approach. The thermal conductivities have been calculated numerically and the power law dependence as the function of temperature has been found. In contrary to hard-KG lattices, all the thermal conductivities of soft-KG lattices increase with temperature monotonically. All these numerical results are in good agreement with the prediction of effective phonon theory. In particular, the exponents of the temperature dependence of thermal conductivities have been found to be quantitatively consistent with the theoretical predictions from effective phonon theory.

\section{acknowledgments}
The numerical calculations were carried out at Shanghai Supercomputer Center, which has been supported by the NSF China with grant No. 11334007 (B.L.). This work has been supported by the NSF China with grant No. 11334007 (L.Y., N.L., B.L.), the NSF China with Grant No. 11205114 (N.L.), the Program for New Century Excellent Talents of the Ministry of Education of China with Grant No. NCET-12-0409 (N.L.) and the Shanghai Rising-Star Program with grant No. 13QA1403600 (N.L.).

\bibliographystyle{apsrev4-1}

\end{document}